\documentclass[useAMS,usegraphicx,usenatbib,referee]{mn2e}
\usepackage{times}
\usepackage{graphicx}
\newcommand{\kms}{\,km\,s$^{-1}$}                                       

\title[V745\,Cassiopeia: an interacting young massive binary in a multiple star system]
      {V745\,Cassiopeia: an interacting young massive binary in a multiple star system\thanks{Based on the data obtained at T\"{U}B\.{I}TAK National Observatory}}

\author[ \"{O}. \c{C}ak{\i}rl{\i} et al.]
  {  \"{O}. \c{C}ak{\i}rl{\i}$^1$\thanks{e-mail:omur.cakirli@gmail.com}, C.~Ibanoglu$^1$, E.~Sipahi$^1$,\\ \\
  $^1$Ege University, Science Faculty, Astronomy and Space Sciences Dept., 35100 Bornova, \.{I}zmir, Turkey\\ 
  }

\begin{document} \maketitle 
\begin{abstract}
We present spectroscopic observations of the massive early type system V745\,Cas, embedded in a 
multiple star system. The brightest star of the system is the eclipsing binary V745\,Cas with an 
orbital period of 1.41 days. The radial velocities of both components and light curves obtained by 
$INTEGRAL$ and $Hipparcos$ missions were analysed. The components of V745\,Cas are shown to be a 
B0\,V primary with a mass M$_p$=18.31$\pm$0.51 M$_{\odot}$ and radius R$_p$=6.94$\pm$0.07 R$_{\odot}$ 
and a B(1-2)\,V secondary with a mass M$_s$=10.47$\pm$0.28 M$_{\odot}$ and radius 
R$_s$=5.35$\pm$0.05 R$_{\odot}$. Our analysis shows that both components fill their corresponding 
$Roche$ lobes, indicating double contact configuration. Using the UBVJHK magnitudes and 
interstellar absorption we estimated the mean distance to the system as 1700$\pm$50\,pc. The locations of the component 
stars in the mass-luminosity, mass-radius, effective temperature-mass and surface gravity-mass are in agreement with those of the 
main-sequence massive stars. We also obtained $UBV$ photometry of the three visual companions and we estimate that all are 
B-type stars based upon their de-reddened colours.We suspect that this multiple system is probably a member of the Cas OB4 
association in the $Perseus$ arm of the $Galaxy$.  
\end{abstract}


 \section{Introduction}\label{sec:intro}
The number of eclipsing massive binary stars, having M $\textgreater$ 9 M$_{\odot}$, is small in 
comparison to lower mass systems (\citet{Gie03}, \citet{Hil05}, \citet{Mas12},   \citet{Moe13}, \citet{San13}). Massive 
stars are usually formed in OB associations. It is well known that massive stars evolve faster and are of interest 
as progenitors of neutron stars and black holes. The evolutionary paths of massive binaries depend on processes 
related to mass transfer between the components and mass loss via stellar wind from one or both components. Consequently, 
in order to better understand the physics of binary systems and  test the predictions of theoretical models, it is thus 
important to quantitatively analyse the properties of massive binary systems with well-constrained orbital parameters. In 
this context studies of the rare early B-type massive stars has major highlights for understanding formation mechanism and 
evolution of massive binaries. \citet{Tor10} collected accurate fundamental parameters, for the 95 double-lined detached 
binary systems. The number of binary systems with masses and radii determined to an accuracy of better than 3\% 
decreases towards the massive stars. The number of systems with at least one component more massive than 
9 M$_{\odot}$ does not exceed ten \citep{Tor10}. Therefore, we initiated a spectroscopic study for the close 
binary systems which include high mass stars located in the upper left-hand corner of the Hertzsprung-Russell diagram.

V745\,Cas (HD1810, HIP\,1805, WDS\,J00229+6214AB, V=8.11, B-V=0.06) is the brightest star of a multiple star 
system. WDS\,J00229+6214 consists of five stars. The angular distances of B, C, D and E from the brightest 
star A are given in the Washington Double Stars Catalog \footnote{ \bf http://ad.usno.navy.mil/wds/} as 9.5, 23.2, 44.9 
and 57.8\,arcseconds, respectively. The visual apparent magnitudes of the stars A, B, C, D, and E are also estimated 
as 8.12, 11.13, 10.8, 12.14 and 12.85\,mag. Despite its relative brightness, V745\,Cas has not been comprehensively 
studied yet. The first spectral classification for V745\,Cas (B0\,IV) has been dating back to \citet{Mor53}. \citet{Hau70} 
gives V=8$^m$.19, (B-V)=0$^m$.08, while \citet{Gue74} gives V=8$^m$.16, (U-B)=-0$^m$.83 and (B-V)=0$^m$.05. Its radial 
velocity, averaging V=-50 \kms, was found to be variable by \citet{San38}, \citet{Wil53}, \citet{Pet61}, and \citet{Eva67}. 
V745\,Cas is listed in the optical pairs by \citet{Mei68} who estimated the spectral types of the A and B components as 
B0\,IVp and A9\,V with a magnitude difference of 1.5\,mag. \citet{Ree03} estimates visual magnitude and color indices of 
the A component as 8$^m$.19, (U-B)=-0$^m$.79 and (B-V)=0$^m$.08.

The light variability of V745\,Cas has been discovered thanks to the Hipparcos satellite mission \citep{Per97}. The 
H$_p$ magnitudes at the maximum and minimum light of the brightest component A of the multiple system are given as 
8$^m$.063 and 8$^m$.164, i.e. a magnitude variation by about 0$^m$.10. The light curve was classified as a W UMa-type 
eclipsing binary according to the criteria of the General Catalogue of variable Stars, with an orbital period of 1.41057 
days from visual inspection of light variation. The Hipparcos' light curve shows that the depths of the eclipses are 
nearly identical. \citet{Kaz99} designated it as V745\,Cas in the 74th Name-list of Variable Stars, and classified 
it as an EB.  

V745\,Cas is located in the sky in the vicinity of the Cas\,OB4 and Cas\,OB14 associations. While the star was not included 
as a member of Cas\,OB4 association by \citet{Hum78} and \citet{Gar92}, it is included into the list of members of Cas\,OB4 
by \citet{Mel95} and \citet{Mel09}. These associations are located in the Perseus arm of the Galaxy. Many estimates have 
been made about their distances to the Sun. These estimates indicate that the Cas OB14 association is closer to the sun 
having a distance of about 1\,kpc, which is less than half the distance of Cas\,OB4. 

This paper is organized as follows. We present new spectroscopic observations and radial velocities of both components 
of the eclipsing pair. By analysing the $INTEGRAL$ and Hipparcos missions' light curves and the new radial velocities 
we obtain orbital parameters for the stars. Combining the results of these analyses we obtain absolute physical 
parameters of both components. In addition, we conclude with a discussion of the system's evolutionary status.

\section{Spectroscopic observations and data reductions}                              
Optical spectroscopic observations of V745\,Cas were obtained with the Turkish Faint Object Spectrograph Camera 
(TFOSC) attached to the 1.5\,m telescope from
January 2 to August 28, 2012, under good seeing conditions. Further details on the telescope and the spectrograph 
can be found at http://www.tug.tubitak.gov.tr. The wavelength coverage of each spectrum was 4000-9000 \AA~in 12 orders, 
with a resolving power of $\lambda$/$\Delta \lambda$ $\sim$7\,000 at 6563 \AA~and an average signal-to-noise ratio 
(S/N) was $\sim$120. We also obtained high S/N spectra of early type standard stars 1\,Cas (B0.5\,IV), HR\,153 (B2\,IV), 
$\tau$\,Her (B5\,IV), 21\,Peg (B9.5\,V), and $\alpha$\,Lyr (A0\,V) for use as templates in derivation of the radial 
velocities. 

The data reduction was performed using the echelle task of IRAF {\footnote{IRAF is distributed by the $National~Optical~
Astronomy$ Observatory, which is operated by the Association of Universities for Research in Astronomy,Inc. (AURA), under 
cooperative agreement with the National Science Foundation} \sc echelle package} \citep{Sim74} following the standard steps:
background subtraction, division by a flat-field spectrum given by a halogen lamp, wavelength calibration using the emission
lines of a Fe-Ar lamp, and normalization to the continuum through a polynomial fit. Heliocentric corrections were computed using
the IRAF {\sc rvsao.bcvcorr} routine and were taken into account in the subsequent radial velocity determination. 

\subsection{Radial Velocities}
Radial velocities of the components were measured by cross-correlation with the {\sc fxcor} task in IRAF 
(\citet{Sim74}, \citet{Ton79}). The standard stars with known radial velocities, given in previous section, were 
used as templates. The standard stars' spectra were synthetically broadened by convolution with the broadening 
function of \citet{Gra92}. The cross-correlation with the standard star 1\,Cas gave the best result. The spectra 
showed two distinct cross-correlation peaks at the quadratures, one for each component of the binary. Since the 
two peaks appear blended, a double $Gaussian$ fit was applied to the combined profile using {\it de-blend} 
function in the task.

The cross-correlation technique was applied to four wavelength regions with well-defined absorption lines 
of the primary and secondary components. These regions include He\,{\sc i} lines at $\lambda$4471 (in the 9th 
order), $\lambda$5876 (in the 3rd order), $\lambda$6678 (in the 3rd order), and $\lambda$7065 (in the 2nd order) 
dominant in early B-type stars. Here we used as weights the inverse of the variance of the radial velocity 
measurements in each order, as reported by {\sc fxcor}. We have been able to measure radial velocities of both 
components with a precision better than 10\,\kms extracted from the {\sc fxcor}. The $Balmer$
lines were not used in the measurements of radial velocities due to their extremely Stark-broadened
and rotationally broadened profiles.

The heliocentric radial velocities for the primary (V$_p$) and the secondary (V$_s$) components are listed in 
Table\,1, along with the dates of observations and the corresponding orbital phases computed with the new 
ephemeris given in Sect. 2.1. The velocities in this table have been corrected to the heliocentric 
reference system by adopting a radial velocity value for the template stars. The radial velocities are 
plotted against the orbital phase in Fig.\,1 where the filled squares correspond to the primary and the 
empty squares to the secondary star. Examination of the $Integral-OMC$ and $Hipparcos$ light curves 
shows no evidence for any eccentricity in the orbit of the system. Therefore, we have started with a circular 
orbit and analysed the RVs using the {\sc RVSIM} software program \citep{Kan07}. The results of the common 
center-of-velocity analysis of the radial velocities are  presented in Table\,2. The best fit was obtained 
for a circular orbit with a mass-ratio of $q=\frac{M_2}{M_1}$=0.571$\pm$0.010. The dotted lines in 
Fig.\,1 show the computed curves. 

\begin{figure*}
\center
\includegraphics[width=16cm,angle=0]{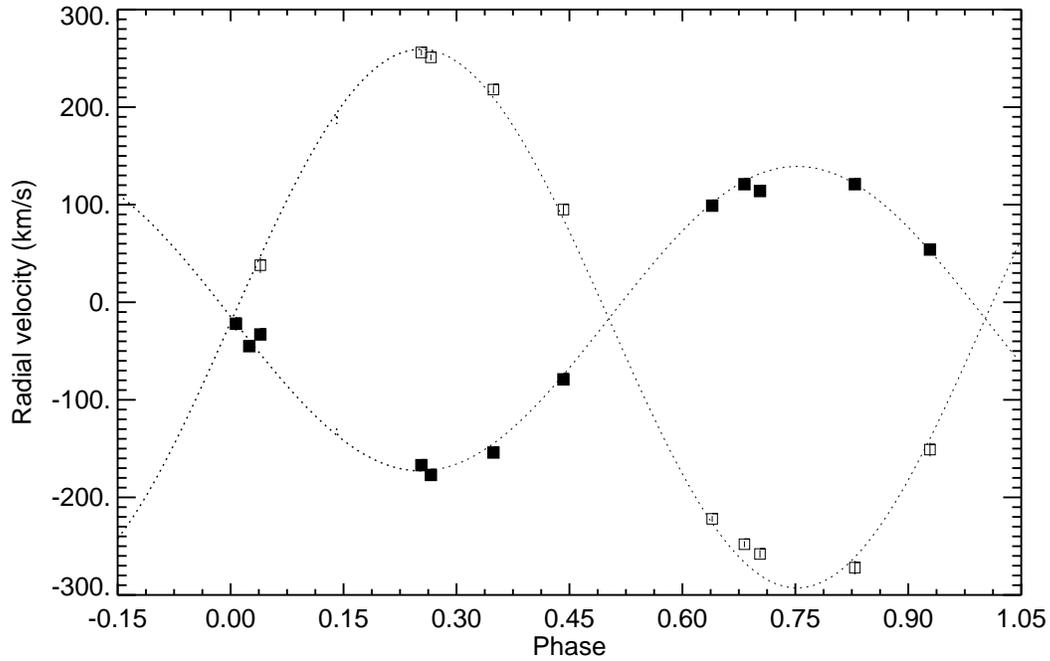}
\caption{Radial velocities for the components of V745\,Cas. Symbols with error bars (masked by the symbol size) 
show the radial velocity measurements for the components of the system (primary: filled squares, 
secondary: open squares).} \end{figure*}

\begin{table}
\scriptsize
\centering
\begin{minipage}{85mm}
\caption{Heliocentric radial velocities of V745\,Cas. The columns give the heliocentric 
Julian date, orbital phase and the radial velocities of the two components with the corresponding 
standard deviations.}

\begin{tabular}{@{}ccccccccc@{}c}
\hline
HJD 2400000+ & Phase & \multicolumn{2}{c}{Star 1 }& \multicolumn{2}{c}{Star 2 } 	\\
             &       & $V_p$                      & $\sigma$                    & $V_s$   	& $\sigma$	\\
\hline
55929.4430  &   0.4420	&  -79  &  5 &     95 &  6 \\
55930.2400  &   0.0071	&  -22  &  7 &	  $--$ &  9 \\
56130.5869  &   0.0395	&  -33  &  6 &	   38 &  8 \\
56131.5227  &   0.7030	&  114  &  3 &	 -258 &  4 \\
56132.4341  &   0.3491	& -154  &  4 &	  218 &  4 \\
56133.3873  &   0.0248	&  -45  &  5 &	  $--$ &  6 \\
56134.5216  &   0.8290	&  121  &  4 &	 -272 &  7 \\
56136.5308  &   0.2534	& -167  &  2 &	  256 &  3 \\
56137.4836  &   0.9288	&   54  &  6 &	 -151 &  7 \\
56162.4659  &   0.6396	&   99  &  2 &	 -222 &  3 \\
56162.5262  &   0.6824	&  121  &  2 &	 -248 &  3 \\
56167.5812  &   0.2660	& -177  &  3 &	  251 &  3 \\

\hline
\end{tabular}
\end{minipage}
\end{table}

\begin{table}
\scriptsize
  \caption { Results of the radial velocity analysis for V745\,Cas}
  \label{parameters}
  \begin{tabular}{lrr}
  \hline
   Parameter 					& Primary	&	Secondary		     \\   
   \hline
  K (km s$^{-1}$) 			  & 156$\pm$2 	      &273$\pm$3		    \\
    V$_\gamma$(km s$^{-1}$) 		 &-16$\pm$2  	      &-16$\pm$2		    \\   
   Average O-C (km s$^{-1}$)		& 4         	      &3      	        \\
   $M$\,$\sin^3 i$ (M$_{\odot}$)  &7.34$\pm$0.20      &4.20$\pm$0.11\\
    $a$ sin$i$ (R$_{\odot}$)		& 4.348$\pm$0.056	  & 7.608$\pm$0.083	\\  
\hline  
  \end{tabular}
\end{table}

\section{Spectral analysis}

Fundamental stellar parameters, such as projected rotational velocity ($v\,sin\,i$), spectral type ($Sp$),
luminosity class, effective temperature (T$_{eff}$), surface gravity ($log~g$), and metallicity ([Fe/H]) may be
determined from the mid-resolution optical spectroscopy.

\subsection{The projected rotational velocities}
Projected rotational velocities of early-type stars were determined usually by means of various methods: the full-width-at-half-maximum
(FWHM) method \citep{Abt02}, the goodness-of-fit ($GOF$) method \citep{Con77}, the cross-correlation-function ($CCF$)
method \citep{Pen96}, the Fourier transform ($FT$) method \citep{Gra76},  and the $iacob-broad$ method \citep{Sim14},
based on a combination of GOF and FT methodologies.

We used the $CCF$-method for measurement of the individual
projected rotational velocities ($v \sin i$) of the component stars. The rotational velocities of the components
were obtained by measuring the full-width-at-half-maximum (FWHM) of the CCFs in five high-S/N spectra of the components  
acquired close to the quadratures, where the spectral lines have the largest Doppler-shift.
The He\,{\sc i} lines at $\lambda\lambda$4471, 6678 and 7065 lines were selected. 
The CCFs were used for the determination of $v\,sin\,i$ through a calibration of the FWHM of 
the CCF peak as a function of the $v\,sin\,i$ of artificially broadened spectra of slowly rotating standard 
star (21\,Peg, $v \sin i$ $\simeq$14 km s$^{-1}$, e.g., \citet{Roy02}) acquired with the same set up and in 
the same observing night as the target. The limb darkening coefficient was fixed at the theoretically
predicted values of 0.42 for both stars \citep{Van93}.  We calibrated the relationship between the CCF Gaussian
width and $v\,sin\,i$ using the \citet{Con77} data sample. This analysis yielded projected rotational
velocities for the components of V745\,Cas as  $V_psin~i$=171 km s$^{-1}$, and $V_ssin~i$=151 km s$^{-1}$. The
mean deviations were 4 and 8 km s$^{-1}$, for the primary and secondary, respectively, between the measured
velocities for different lines.

\subsection{The spectral classification}
Spectral types of the components were first estimated by comparison both with our standard stars' spectra and 
also with templates taken from the \citet{Val04} $Indo-U.S.\ Library\ of\ Coude\ Feed\ Stellar\ Spectra$ (with 
a resolving power of about R=3600) that are representative of stars with various metallicities, spectral types 
from late-O type to early-A, and luminosity classes V, IV, and III.

We have performed a spectral classification for the components of the system using COMPO2, an IDL ({\sc Interactive 
Data Language RSI}) code for the analysis of the spectra of SB2 systems written by \citet{Fra06}. This code, similar 
to the {\sf SYNTHE} code \citep{Kur81}, was adapted to the TFOSC spectra of the binary systems. This code searches 
for the best combination of two standard-star spectra able to reproduce the observed spectrum of the system. We give, 
as input parameters, the radial velocities and projected rotational velocities $v\sin i$ of the two components, which 
were already derived. The code then finds, for the selected spectral region, the spectral types and fractional flux 
contributions that better reproduce the observed spectrum, i.e. which minimize the residuals in the collection of 
difference (observed\,$-$\,constructed) spectra.

The atmospheric parameters of the reference stars given by \citet{Val04} were recently revised by \citet{Wu11}. For 
this task we selected about 200 single-star' spectra spanning the ranges of expected atmospheric  
parameters, which means that we have searched for the best combination of spectra among 39204 possibilities per 
each spectrum. 

The observed spectra of V745\,Cas around the $\lambda\lambda$4471, and 6678 spectral lines were best represented by 
combination of the spectra of HD\,34816 (B0.5 V, $log~g$=3.91) and HD\,184915 (B1.5 III, $log~g$=3.40). We have derived the 
spectral types for the primary and secondary component of V745\,Cas as B(0$\pm$0.5)\,V and B(1.5$\pm$0.5)\,III, 
respectively. The atmospheric parameters obtained by the code are presented in Table\,3. The observed spectra of 
V745\,Cas at nearly quadratures around the He\,{\sc i} lines at $\lambda\lambda$4471, and 6678 are compared with 
the combination of two  single-star' spectra in Fig.\,2.

This result can be confirmed using the ratios of equivalent widths of He {\sc i } 4471 and He {\sc ii} 4541 lines. Spectral 
lines of the components are well separated at the quadratures of the eclipsing pair. We measured equivalent widths (EW) 
of the primary star He {\sc i } 4471 and He {\sc ii} 4541 lines for estimating {\bf its spectral type}. The average 
logarithmic ratio of the equivalent widths (EW) of the primary is about 1.19$\pm$0.17, which corresponds to a B0 
type star \citep{Huc96}.  The luminosity criterion based on the logarithm of the ratio of the line strengths of 
Si\,{\sc iv} 4088 and He\,{\sc i} 4143 \citep[]{Con86,Huc96} yields $-1.07 \pm 0.05$. This value corresponds 
to main-sequence (luminosity class V) stars.

Further support for this classification comes from the photometry of V745\,Cas available in the literature. The 
visual apparent magnitude and colour indices were given by \citet{Gue74} as $V$=8$^m$.16, ($U-B$)=$-$0$^m$.83 
and ($B-V$)=0$^m$.05 with an uncertainty of $\pm$0.01 mag. The quantity $Q$=($U-B$)$-$($E_{(U-B)}$/($E_{(B-V)}$)$(B-V)$ 
is independent of interstellar extinction. The average value of the ratio ($E_{(U-B)}$/($E_{(B-V)}$) is 0.72$\pm$0.03 
(\cite{Joh53}; \cite{Hov04}). We compute the reddening-free index from the data of \citet{Gue74} as $Q$=$-$0.866$\pm$0.014. 
The $Q$ $-$ values were calculated by \citet{Hov04} begining from O8 to G2 spectral types for the luminosity classes 
between main-sequence and supergiants. The $Q$$-$$Sp$ calibration yields an approximate spectral type of the system 
as B0 with a range from O9 to B1. The spectral classification from the spectra and photometric indices are quite 
consistent. 

In addition, infrared colors J$-$H=$-$0.057$\pm$0.064, H$-$K=0.015$\pm$0.062\,mag are given in the 2MASS 
catalog \citep{Cut03}. These indices confirm spectral classifications made by spectra and wide-band 
photomeric indices. A preliminary analysis of the light curve (see Section 4.3) yields light ratio of 0.43 for the V-passband.
This light ratio, the observed colours and the intrinsic colour of the primary star of $(B-V)_0$=$-$0.30$\pm$0.01 \citep{Dri00} allow us to estimate
an intrinsic composite colour of $(B-V)_0$=$-$0.29$\pm$0.01. Thus, the interstellar 
reddening of $E_{(B-V)}$=0.34$\pm$0.01\,mag is estimated for the system.

\begin{table}
\scriptsize
\centering
\begin{minipage}{85mm}
\caption {Spectral types, effective temperatures, surface gravities, and rotational velocities of each star 
estimated from the spectra of V745\,Cas.}
\begin{tabular}{@{}lcccccccc@{}c}
\hline
Parameter    			& \multicolumn{2}{c}{\sf V745\,Cas}	&  			\\
             			& Primary                   		& Secondary   		\\
\hline
 Spectral type 			& B(0$\pm$0.5)\,V 			&B(1.5$\pm$0.5)\,III	\\
 T$_{eff}$ (K)	    		&30\,000$\pm$880  			&25\,600$\pm$1\,050	\\   
 $\log~g$ ($cgs$)		& 3.91$\pm$0.00         		&3.40$\pm$0.00       	\\    
 $Vsin~i$ (km s$^{-1}$)  	&171$\pm$4	  		&151$\pm$8     	\\  
\hline
\end{tabular}
\end{minipage}
\end{table}

\begin{figure*}
\center
\includegraphics[width=14cm,angle=0]{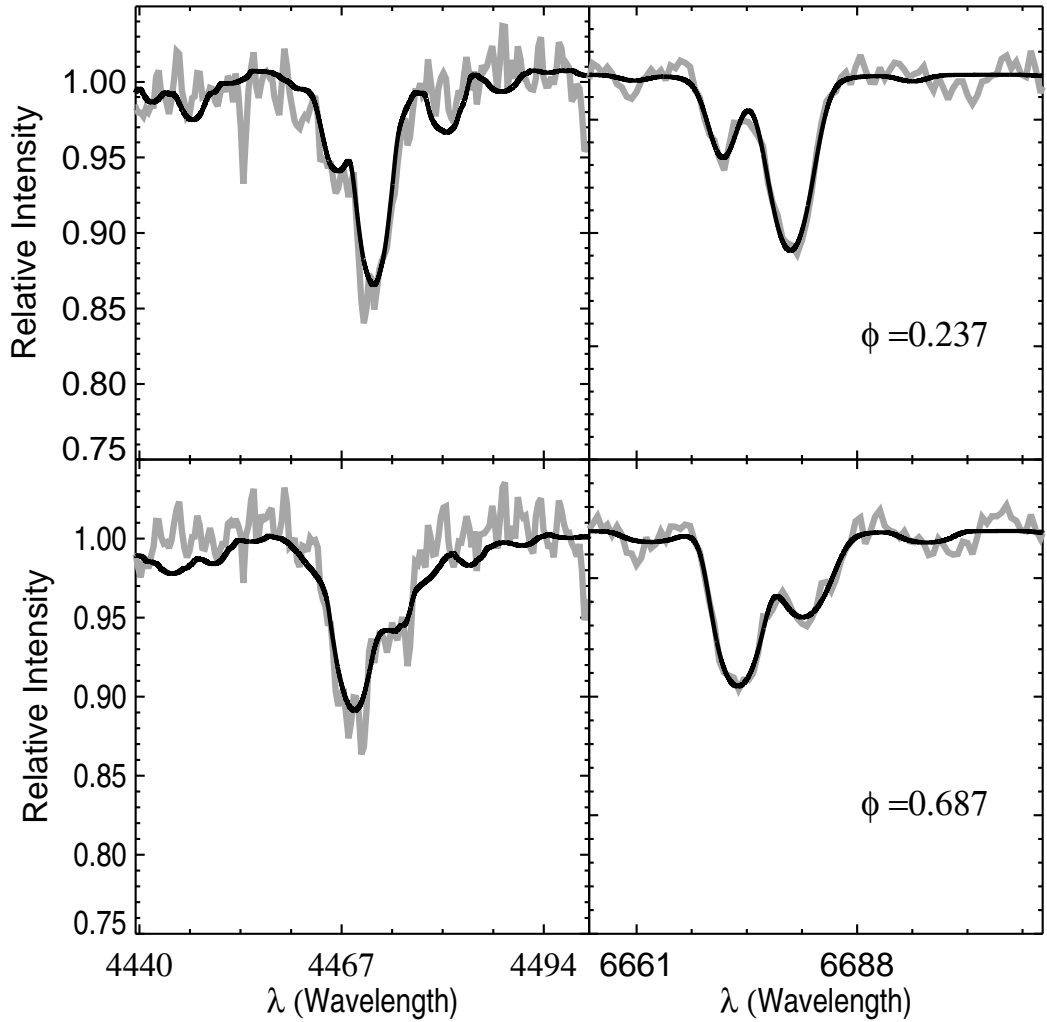}
\caption{Comparison between the observed spectra of V745\,Cas obtained near quadratures and the best-fitting 
spectra around He\,{\sc i} $\lambda$4471 (left panel) and  the  $\lambda$6678 lines (right panel). The deeper 
lines indicate the primary and the shallow ones the secondary star.}
\end{figure*}

\section{Analyses of the light curves}
\subsection{Light Curve Constraints}
The first photometric observations of V745\,Cas were made by the Hipparcos mission and 144 H$_p$ magnitudes were 
listed by \citet{Van07}. These magnitudes were obtained in a time interval of about three years. The average accuracy 
of the Hipparcos data was given as $\sigma_{H_p}$ $\sim$ 0.01\,mag. The light variation from peak-to-peak is about 
0.12\,mag. However, we have estimated a scatter of about 0.022\,mag at the maxima and within the primary minimum.
The scatter at the secondary eclipse is slightly larger, amounting to 0.028\,mag.  

The second set of photometric observations were made by the $International~Gamma-Ray~Astrophysics~Laboratory$ ($INTEGRAL$)
mission \citep{Alf13}]. The $Integral-OMC$ database contains light curves for many previously unknown variable stars.
We extracted the V-passband light curve for V745\,Cas ($Integral-OMC$\,4019000053) from this catalog. The $INTEGRAL$
data are consisting of 1\,980 Johnson's V-passband magnitudes. \citet{Dom03} estimate an average accuracy of about
0.006\,mag for objects of V=12\,mag.

The 1\,980 photometric measurements of $Integral$, including eclipses, permit determination of the
time of mid-primary eclipse and the orbital period of the system. A periodogram analysis, e.g. PERIOD04 
\citep{Len05}, has been applied to the data obtained by $INTEGRAL$ mission. We determine following ephemeris 

\begin{equation}
Min I(HJD)=2\,455\,222.1234(2)+1^d.410571(7) \times E
\end{equation}
where the standard deviations in the last significant digits are given in parentheses. However, \citet{Alf12}
determined an orbital period of 1.4106019\,d, which is slightly longer than we found. Periodogram analysis
indicates that there is a light variation with a second period of about 0.1659$\pm$0.0007 days (approximately 4 h)
having a peak-to-peak amplitude of about 0.014\,mag. If this light variation really exists it may originate from 
intrinsic variation of one or both stars. The spectral types of the components are similar to those of $\beta$\,Cephei 
type pulsating stars. Therefore such intrinsic variation may be expected from the components of V745\,Cas. However, 
existence of such a variation should be checked by precise multi-passband photometric observations.   

We have measured the magnitude difference between the A and B components of the visual pair as 3.13\,mag 
(see Chapter 4) in the V-passband. This difference is slightly larger than the difference of 3.01\,mag given by 
the $WDS~catalogue$. Since the INTEGRAL Optical Monitoring Camera has an angular resolution of about 23\,arcsec 
its photometric  data cover the light contribution of the component B. Light contribution of the component B to the total 
light has been calculated as 0.053 which slightly affects the observed magnitudes. Therefore we subtracted the 
light contribution of component B from all the measured magnitudes. On the other hand the $H_p$ magnitudes measured
by the $Hipparcos$ mission were transformed to the Johnson's V-passband using the transformation coefficients given
by \citet{Har98}. Then both data sets were combined for a simultaneous analysis. All available photometric data are 
phased and plotted in Fig.\,3, where the open circles represent the Hipparcos data. It is clear from Fig.\,3 that 
V745\,Cas is an eclipsing binary, with a W\,UMa type light curve that is dominated by variations due to tidal distortion. 

In order to model the photometric light curve, we have used the computer programme $PHOEBE$ (PHysics Of Eclipsing 
BinariEs) developed by \citet{Prs05}. This program is an implementation of the Wilson-Devinney (hereafter WD) 
binary star code \citep{Wil71}. The Roche geometry, sensitive to the mass-ratio, is used to represent the tidal 
deformation of the two stars. One of the main difficulties in this modelling is determination of effective 
temperature of the primary star and the mass-ratio of the system. Effective temperatures of both stars could 
already be determined from the spectra. In addition the mass-ratio, which is the second key-parameter for 
the modelling, was also obtained from the radial velocities of the stars.

\begin{figure*}
\center
\includegraphics[width=14cm,angle=0]{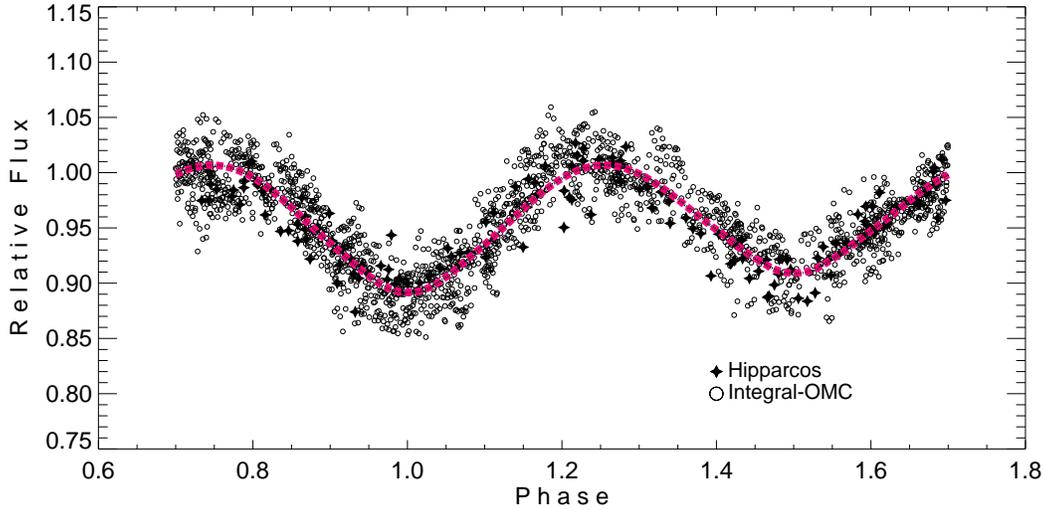}
\caption{V-passbad light curve constructed from the $INTEGRAL$ (empty circles) and $Hipparcos$ 
(filled stars) photometric data of V745\,Cas. The continuous and dashed lines show the best-fit 
models obtained by the WD code using the Mode 6 and Mode 3, respectively.} \end{figure*}

\begin{table}
\scriptsize
\caption{Results of the analyses of the light curve for V745\,Cas.}
\begin{tabular}{lrrrrr}
\hline
Parameter & Mode 2 & Mode 3 & Mode 4 & Mode 5 & Mode 6  \\
\hline	
$i^{o}$			        &43.65$\pm$0.03		&47.30$\pm$0.32	&44.08$\pm$0.12	&48.07$\pm$0.20	    &47.51$\pm$0.22      \\
T$_{eff_1}$ (K)			&30\,000[Fix]		&30\,000[Fix]		&30\,000[Fix]		&30\,000[Fix]	    &30\,000[Fix]        \\
T$_{eff_2}$ (K)			&14\,210$\pm$235	&25\,350$\pm$315	&17\,580$\pm$320	&25\,350$\pm$301    &25\,540$\pm$300     \\
$\Omega_1$			&3.0572$\pm$0.0079	&3.0098$\pm$0.0036	&3.0099       		&3.0181$\pm$0.0067    &3.0099$\pm$0.019     \\
$\Omega_2$			&2.8105$\pm$0.0040	&3.0080$\pm$0.019	&2.8908$\pm$0.0081	&3.0099$\pm$0.019    &3.0099$\pm$0.019     \\
$r_1$				&0.4202$\pm$0.0015	&0.4285$\pm$0.0007	&0.4281$\pm$0.0016	&0.4251$\pm$0.0013  &0.4281$\pm$0.0016   \\
$r_2$				&0.3816$\pm$0.0012	&0.3301$\pm$0.0007	&0.3572$\pm$0.0021	&0.3297$\pm$0.0015  &0.3297$\pm$0.0015  \\
$\frac{L_{1}}{(L_{1}+L_{2})}$   &0.8274$\pm$0.0033	&0.7033$\pm$0.0047	&0.7951$\pm$0.0048	&0.7002$\pm$0.0047  &0.6979$\pm$0.0046 \\
$\sum(O-C)^{2}$			&1.0242			&1.0484              	&1.0353              	&1.0486              &1.0477              \\		
\hline
\end{tabular}
\end{table}

\begin{figure*}
\center
\includegraphics[width=14cm,angle=0]{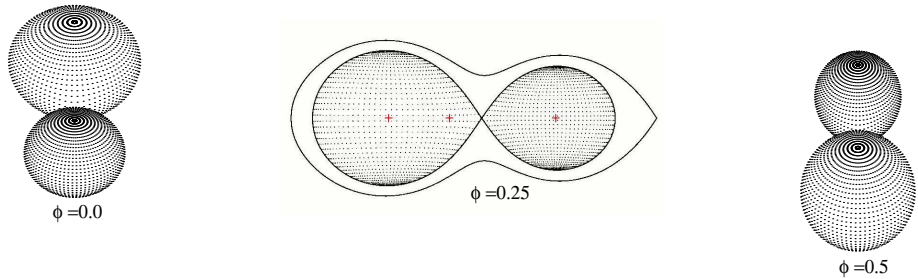}
\caption{Equipotential structure of V745\,Cas corresponding to the solution of the light curve using Mode 6 at the mid-primary (left) and 
mid-secondary  (right) eclipses. Comparison with the Roche lobes is shown in the middle.  } \end{figure*}

Linear limb-darkening coefficients were interpolated from the tables of \citet{Van93}. They are 
updated at every iteration by $PHOEBE$. The gravity-brightening coefficients $g_1$=$g_2$=1.0 and albedos 
$A_1$=$A_2$=1.0 were fixed for both components, as appropriate for stars with radiative atmospheres.

\subsection{Light Curve Solutions}

We started with the  $Mode-2$ of the Wilson-Devinney code, referring to the detached Algols for the analysis
of light curves as described by \citet{Wil06}. The adjustable parameters in the light curve fitting were the orbital
inclination ($i$), the effective temperature of the secondary star (T$_{eff_2}$), the potentials ($\Omega_1$ and $\Omega_2$), 
the luminosity of the primary (L$_1$), and the zero-epoch offset. The parameters of the solution are tabulated as $Mode-2$ in 
Table\,4. The effective temperature, fractional luminosity and fractional radius of the secondary are about 14 000 K, 0.17, and 
0.38, respectively. The effective temperature and fractional luminosity are too small, nearly half, when compared to the values 
estimated from the spectra. We obtain an inclination of about 43.7 degrees which yields masses of about 22 and 13 M$_{\odot}$ 
for the primary and secondary star respectively. When the volumes of the stars are compared with their corresponding Roche lobes we see that 
while the primary fills up its lobe the secondary star overflows its lobe. Since the configuration is inconsistent with the 
detached assumption the results of the analysis are taken to be unacceptable. 
 
Next, we tried $Mode-3$ (for overcontact systems, the stars are in geometrical contact without being in thermal contact).
The adjustable parameters were $i$, T$_{eff_2}$, and L$_1$. The results of the analysis are presented in the third column 
of Table\,4. Analysis with this mode yields higher $i$, T$_{eff_2}$, and L$_2$ but lower fractional radius for the secondary star.
Although the sum of residuals squared is slightly larger from that obtained in $Mode-2$ these results are in agreement with those 
obtained from the spectra. However, comparison with the Roche lobes shows that both components are in contact with their inner Roche lobes,
which is inconsistent with the assumption of overcontact configuration. In other words, the solution clearly reveals a fill-out factor 
$f$=($\Omega_{in}$ $-$ $\Omega$)/($\Omega_{in}$ $-$ $\Omega_{out}$)=0, corresponding to the contact configuration.     

The results of the analyses using $Mode-4$ (primary star fills its lobe) and $Mode-5$ (secondary star fills its lobe) were given
in the fourth and fifth columns of Table\,4. The analysis with $Mode-4$ gives similar results as $Mode-2$, but slightly larger 
$\sum(O-C)^{2}$. While the primary component fills its lobe the secondary overfills its lobe which is inconsistent with the 
$a$ $priori$ assumption. The analysis under the assumption of $Mode-5$ gives the parameters which are very close to those 
obtained by the $Mode-3$ solution. However, comparison with the Roche lobes shows that both components are just in contact 
with their corresponding Roche lobes.     

Comparison with the Roche lobes
points out double contact configuration. Therefore, we applied $Mode-6$ (for double contact systems) for the analysis of the
observed light curve. The results of this analysis are given in the last column of Table\,4. The uncertainties assigned to 
the adjusted parameters are the internal errors provided directly by the code. While the sum of residuals squared is
slightly larger than those obtained by $Mode-2$ and $Mode-4$, it is slightly smaller than those obtained by $Mode-3$ and $Mode-5$.
However, it should be noted that the solutions obtained with $Mode-3$, $Mode-5$ and $Mode-6$ are of similar quality and yield similar 
model parameters. The results obtained with $Mode-6$ are in good agreement with the atmospheric parameters determined from the 
spectra. The computed light curve is compared with the observations in Fig.\,3. This solution indicates that the grazing 
eclipse, lasting about five hours, occurs in the V745\,Cas system. In Fig.\,4 we show the equipotential surfaces of both 
components for the light curve solution at the phases mid-primary and mid-secondary eclipses. In the middle of the figure 
the volumes of the stars are compared with the Roche lobes.

\begin{figure*}
\begin{center}
   \includegraphics[width=14cm]{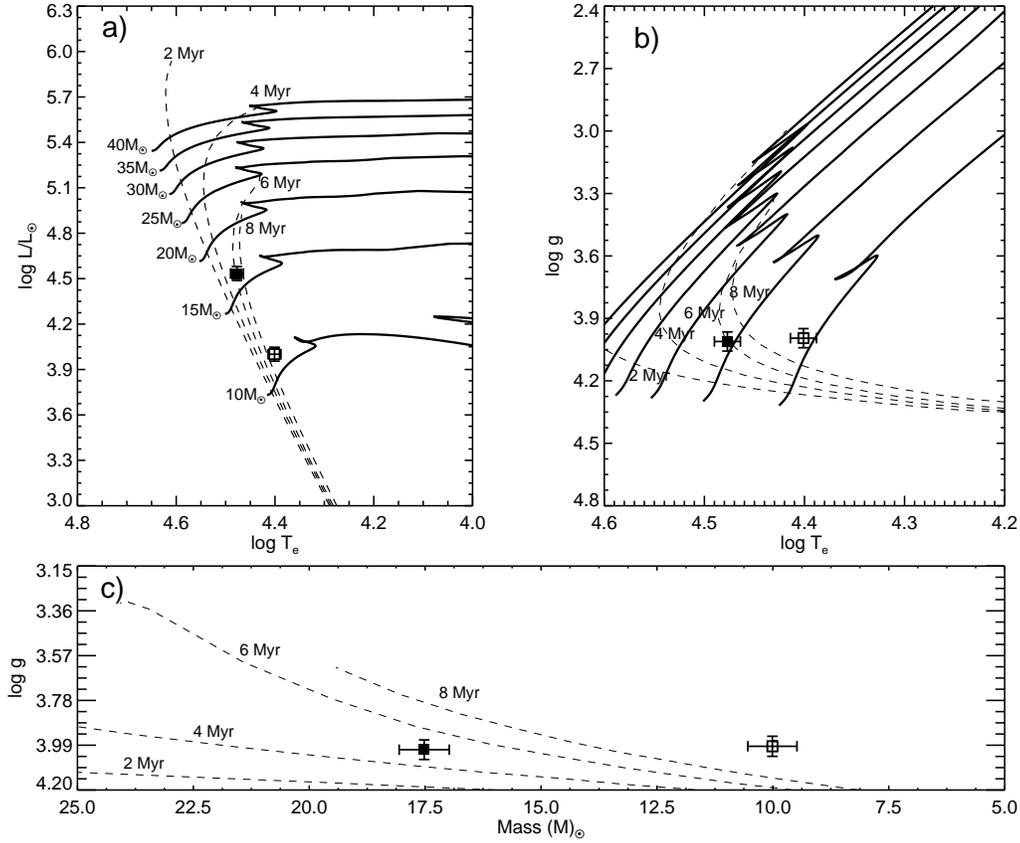}
\end{center}
  \caption{Positions of the components of the system in the luminosity-effective temperature and 
  gravity effective temperature planes. The solid lines show evolutionary tracks for 
  stars with masses of 40, 35, 30, 25, 20, 15 and 10 solar masses taken from \citet{Eks12}. }  \label{fig:evo}
\end{figure*}

\section{Results and discussion}
Combining the results of radial velocities and light curve analyses we have calculated the absolute parameters 
of the stars. Separation between the components of the eclipsing pair is calculated as $a$=16.21$\pm$0.14R$_{\odot}$. 
The fundamental stellar parameters for the components such as masses, radii, luminosities are listed in Table\,5 
together with their formal standard deviations. The standard deviations of the parameters have been determined by 
means of the JKTABSDIM\footnote{This can be obtained from http://http://www.astro.keele.ac.uk/$\sim$jkt/codes.html} code 
\citep{Sou05}. The mass for the primary of $M_p$=18.31$\pm$0.51 M$_{\odot}$ and secondary of 
$M_s$=10.47$\pm$0.28 M$_{\odot}$ are consistent with B0\,V and B(1-2)\,IV-V stars.

The bolometric magnitudes of the components are computed as -6.62$\pm$0.14 and -5.35$\pm$0.06 mag for the primary
and secondary components taking the absolute bolometric magnitude of the Sun as 4.74. The bolometric corrections listed
in Table 5 are taken from  \citet{Lan07} adopting V$_{tur}$= 2 km s$^{-1}$, and $[Z/Z_{\odot}]$=1.0 (solar metallicity)
and the effective temperatures and the surface gravities of the components given in the table. Using the interstellar reddening
of $E$(B-V)=0.34\,mag and absorption coefficient of 3.1 we estimate the absorption in visual band as 1.05 mag. With V=8.18 at
the maximum light one finds $V_{0p}$=7.52 and $V_{0s}$=8.43 for the primary and secondary star, respectively. Then the 
distance modulus are 11.16 and 11.15 mag which correspond to the distances of 1\,703$\pm$63 and 1\,697$\pm$38 pc. As a 
weighted mean, we find the distance to the system as 1\,700$\pm$50\,pc.

Our line-of-sight 
passing through the binary system V745\,Cas traverses first the $Perseus$ arm of the $Galaxy$ and then the distant $Cygnus$ 
arm. The mean residual radial velocity and Galactocentric distance of the OB-associations in the $Perseus$ region, including 
Cas\,OB4, were given as 8.4\,kpc and -6.7km s$^{-1}$. We estimated the distance of V745\,Cas from the galactic plane as 20.4\,pc 
and from galactic center as 8.5\,kpc, where the distance of sun from the galactic plane is taken as 20.5\,pc \citep{Hum95}. The 
distances of the associations in the $Perseus$ arm from the Sun were estimated between 1.8 and 2.8\,kpc \citep{Mel09}. Therefore, 
we suspect that V745\,Cas may be a member of the Cas\,OB4 association. If it is a member it should be located at the nearest 
edge of the association. This conclusion agrees with the result of \cite{Mel95} who quoted V745 Cas as a member of Cas OB4, for 
which they gave a distance of 2.04\,kpc.

\begin{table}
\scriptsize
  \caption{Properties of the V745\,Cas components}
  \label{parameters}
  \begin{tabular}{lrr}
  \hline
   Parameter 						            & Primary	       &	Secondary		      \\   
   \hline
	Mass (M$_{\odot}$)				        & 18.31$\pm$0.51		&10.47$\pm$0.28		\\
	Radius (R$_{\odot}$)				    & 6.94$\pm$0.07		    & 5.35$\pm$0.05		\\   
	$T_{eff}$ (K)					        & 30\,000$\pm$1000	    & 25\,540$\pm$300	\\
   $\log~(L/L_{\odot})$				        & 4.547$\pm$0.056	    & 4.040$\pm$0.025	\\
   $\log~g$ ($cgs$) 					    & 4.018$\pm$0.006 	    & 4.002$\pm$0.007	\\
   $Sp.Type$ 						        & B0V				    & B(1-2)IV-V  		\\
   $M_{bol}$ (mag)					        & -6.62$\pm$0.14		&-5.35$\pm$0.06		\\
   $BC$ (mag)						        & -2.98          	    &-2.63   			    \\  
   $M_{V}$ (mag)						    & -3.64$\pm$0.14		&-2.72$\pm$0.06		\\  
   $(vsin~i)_{calc.}$ (km s$^{-1}$)	        & 184$\pm$2			    & 142$\pm$2			  \\       
   $(vsin~i)_{obs.}$ (km s$^{-1}$)	        & 171$\pm$4		        & 151$\pm$8 		  \\    
   $d$ (pc)							        & 1\,703$\pm$63	 	    & 1\,697$\pm$38		\\
\hline  
  \end{tabular}
\end{table}

The distance to the Cas\,OB4 association had been subjected to various studies. There are a number of well-known OB associations 
and galactic open clusters in the direction of Cas\,OB4 association (approximately $l$=120, $b$=0 degrees). The mean interstellar 
reddening of 0.40, 0.34, 0.55, {\bf 0.47} and 0.55 mag, and distance of 1.43, 2.96, 3.8, 1.67 and 3.47 \,kpc were found for Mayer1 
\citep{Kha05}, King\,14 \citep{Net06}, NGC\,103 \citep{Phe93}, NGC\,129 \citep{Tur92} and NGC\,146 \citep{Sub05}, respectively. The 
derived systemic velocity, reddening and distance of V745\,Cas agree quite well with those values of NGC\,129.

We have observed the components A, B, C and D of the multiple stellar system (WDS\,J00229+6214) simultaneously using the 100\,cm telescope 
of the Turkish National Observatory of Turkey. We used the A-star as a reference for which UBV magnitudes are known. The photometric 
observations were made when the component A was at maximum light. The wide-band UBV magnitudes of the components B, C and D were 
obtained with respect to the component A and are presented in Table\,6. The 
standard deviations of the measurements are about 0.01\,mag. The apparent visual magnitude for C-star with a separation of 23.2\,arcsec  
was given in the WDS catalogue as 10.8 which is inconsistent with that we determined. We suspect that the C-star may be variable in 
brightness. Using the Johnson Q$-$method we determined the spectral types of the stars as B3, B8 and B7 for B, C and D, respectively. 
According to this classification the interstellar reddening for B and C is similar with that of the eclipsing pair. However 
the D star appears to have the largest reddening, amounting to E(B-V)=0.60\,mag. Such a result is expected because the reddening 
may vary from region to region in the association depending on the density of molecular clouds.

\begin{table}
\scriptsize
\caption{UBV measurements of the components B, C and D of the multiple stellar system and their probable spectral types.}
\begin{tabular}{lrrrrr}
\hline
Star & U & B  & V  & Q  & Sp.Type \\
\hline	
$B$			&10.82	&11.32  &11.25 &-0.542 & B3 					\\
$C$			&13.47  &13.54  &13.33 &-0.218 & B8						\\
$D$			&12.62  &12.65  &12.18 &-0.357 & B7						\\			
\hline
\end{tabular}
\end{table}

Figure\,5 shows the components, with 1-$sigma$ error bars, of V745\,Cas in the log T$_{eff}$-log L/L$_{\odot}$ 
(left panel) and log T$_{eff}$-log\,g planes (right panel). The evolutionary tracks and isochrones for the
non-rotating single stars with solar composition are taken from \citet{Eks12}. The models considering mass
loss and convective overshooting are adopted. We typically compare the positions of the stars with evolutionary
tracks and determine what masses they would predict. This comparison points out a mass slightly smaller for
the primary but higher for the secondary. The locations of the components in both
panels show that the components are in the main sequence band, not far from the ZAMS. Although both components are 
young they fill their corresponding Roche lobes due to very small separation. Therefore, one should not necessarily expect
a good agreement with single star tracks.

V745\,Cas is one of a few massive contact binaries, namely TU\,Mus, V382\,Cyg, and LY\,Aur \citep{Pen08}. The masses of 
the components as well as the orbital period are very similar to that of TU\,Mus. As pointed out by \citet{Pen08} the 
observed mass ratios of the contact systems are slightly higher than those of the semi-detached systems. The components 
of the contact systems have similar spectral types (within one spectral type of each other) and luminosity classes. 
\citet{Wel01} presented evolutionary calculations for 74 systems considering case A and B scenarios. Our analyses 
indicate that  V745\,Cas is a contact system. Both components have nearly the same luminosity class and almost 
identical surface gravities, as given in Table\,5. Therefore, the system is most probably undergoing $case~A$ 
evolution. According to the \citet{Wel01} models, $case~A$ contact systems can occur when the initial period of 
the system is very small and/or mass ratio is either very small or very large. The observed properties of the 
components are similar with those of the system No.53, (initial masses of 16 and 12 M$_{\odot}$ and orbital period 
of 1.5 days) in their Table\,3. Their model shows that the more massive star reaches its Roche lobe fairly early 
in its evolution. Rapid mass transfer from the more massive star to the less massive star occurs, the orbital period 
decreases which causes the loser star to overflow at a faster rate. Large increase in mass of the gainer star and 
smaller orbit cause it to expand and reach its $Roche$ lobe. The orbit of the binary continues to shrink until the 
gainer star is more massive. Then the mass transfer after this point causes the orbit to expand, leading to a longer 
orbital period. Fast $case~A$ mass transfer still continues until the loser is much less massive. Since the ratio 
of mass in the core to that of the envelope of the loser is increasing this causes its radius to increase. Finally 
the gainer expands and reaches to the Roche lobe due to the increase in mass. Thus a contact system, similar to 
V745\,Cas, is formed.

\begin{figure*}
  \begin{center}
      \includegraphics[width=12cm]{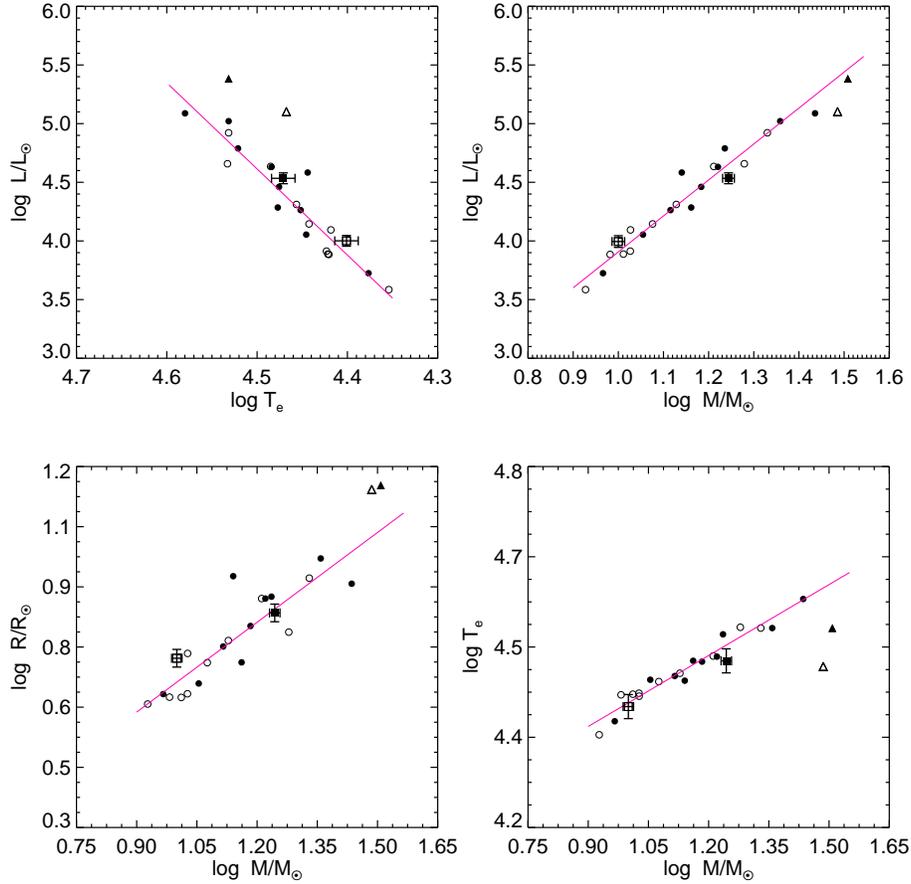}
  \end{center}
  \caption{Locations of the primary (solid dots) and secondary (open circles) components of the detached, young 
  eclipsing binary systems in various diagrams. (a) log L - log T$_{eff}$, (b)log L - log M, (c) log T$_{eff}$ - log M 
  and (d) log R - log M. The solid lines show the best correlations between the parameters. The component of 
  V745\,Cas and MY\,Ser are shown by squares and triangles. }  \label{fig:evo}
\end{figure*}

In Fig.\,6 we show the correlations between luminosity-effective temperature, mass-luminosity, mass-radius and 
mass-effective temperature for the components of the detached massive eclipsing binaries where the filled circles 
refer to the primaries and the open circles to the secondaries. The parameters are taken from \citet{Tor10}. These 
stars lie close to the zero-age main-sequence on the HR diagram. Though the sample contains only ten stars reliable 
correlations between the parameters could be obtained. Linear least-squares fit to the data gives following relationships,

\begin{equation}
log\frac{L}{L_{\odot}}=7.34(\pm0.60) ~~ log T_{eff}-28.48(\pm2.69) \\
\end{equation}

\begin{equation}
log\frac{L}{L_{\odot}}=3.07(\pm0.17) ~~ \frac{M}{M_{\odot}}+0.84(\pm0.19) \\ 
\end{equation}

\begin{equation}
log T_{eff}=0.39(\pm0.02) ~~ log\frac{M}{M_{\odot}}+4.01(\pm0.03)
\end{equation}

\begin{equation}
log\frac{R}{R_{\odot}}=0.75(\pm0.10) ~~ log\frac{M}{M_{\odot}}-0.08(\pm0.11)
\end{equation}

The numbers in the parentheses are the standard errors of the preceding coefficients determined by linear least 
squares solutions. The components of V745\,Cas and MY\,Ser are shown by squares (primary:filled, secondary:empty)  
and triangles (primary:filled, secondary:empty) in Fig.\,5. While the 
parameters for V745\,Cas are determined in this study, the parameters for the components of MY\,Ser are adopted 
from \citet{Iba13}. The positions of the components of V745\,Cas agree well with those of the components of 
detached, young massive eclipsing binaries. This may be taken as an indicator that mass-exchange between the 
components or mass-loss from them did not yet lead to a significant change on their radii and effective temperatures.

\section{Conclusion}

V745\,Cas is one of the rare early type contact systems with massive components. Analysis of the radial 
velocities and light curves yielded absolute parameters of the components. The components are classified 
as B0V and B(1-2)V spectral types with masses of 18.31$\pm$0.51 and 10.47$\pm$0.28 M$_{\odot}$. 
Comparison with the evolutionary models and with the parameters of detached massive binaries show that both 
components are still close to the zero-age main-sequence. Although the components of V745\,Cas  are in contact
with their Roche lobes they locate on the main-sequence band. Using our estimation of $E$(B-V) and A$_v$ we determined
a distance of about 1700$\pm$50\,pc which is in agreement with the distance estimations of the Cas\,OB4 association.
We also observed V745\,Cas, the brightest component of the multiple star system with the other components simultaneously.
These observations indicate that the components B, C and D are also massive stars with spectral types of B3, B8 and B7.

\section*{Acknowledgments}
We thank to T\"{U}B{\.I}TAK National Observatory (TUG) for a partial support in using RTT150 
telescope with project number 11BRTT150-198.
We also thank to the staff of the Bak{\i}rl{\i}tepe observing station for their warm hospitality. This 
study is supported by Turkish Scientific and Technology Council under project number 112T263.
The following internet-based resources were used in research for this paper: the NASA Astrophysics Data 
System; the SIMBAD database operated at CDS, Strasbourg, France; and the ar$\chi$iv scientific paper 
preprint service operated by Cornell University. 
This research was supported by the Scientific Research Projects Coordination Unit of Istanbul University. Project 
number 3685. We thank \c{C}anakkale Onsekiz Mart University Astrophysics Research Center and Ulup{\i}nar Observatory 
together with \.{I}stanbul University Observatory Research and Application Center for their support and allowing use 
of IST60 telescope. The authors thank to the anonymous referee for helpful comments that improved the clarity of the text. 

\bibliographystyle{elsarticle-harv}



\end{document}